\documentstyle[12pt,aasms4,psfig]{article}
\tightenlines

\font\cap=cmcsc10

\def\hi{\noindent \hangindent=2.5em}

\def\kpc{{\rm\,kpc}}

\def\kmsec{{\rm\,km/s}}

\def\surfb{{\rm\,mag/arcsec^2}}

\def\spose#1{\hbox to 0pt{#1\hss}}
\def\lta{\mathrel{\spose{\lower 3pt\hbox{$\mathchar"218$}}
     \raise 2.0pt\hbox{$\mathchar"13C$}}}
\def\gta{\mathrel{\spose{\lower 3pt\hbox{$\mathchar"218$}}
     \raise 2.0pt\hbox{$\mathchar"13E$}}}

\lefthead{Dalcanton \& Shectman}
\righthead{Chain Galaxies Are Edge-On LSBs}
 
\begin{document}

\title{``Chain'' Galaxies are Edge-On Low Surface Brightness Galaxies}

\author{Julianne J.\ Dalcanton\altaffilmark{1,2} \& Stephen A.\ Shectman}
\affil{	Observatories of the Carnegie Institute of Washington\\
	813 Santa Barbara Street, Pasadena CA, 91101}

\altaffiltext{1}{e-mail address: jd@ociw.edu}
\altaffiltext{2}{Hubble Fellow}

\bigskip
\centerline{\it to appear in The Astrophysical Journal Letters}
\bigskip

\begin{abstract}

Deep {\sl HST} WFPC2 images have revealed a population of very narrow
blue galaxies which Cowie et al.\ (1996) have interpreted as being a
new morphological class of intrinsically {\sl linear} star forming
galaxies at $z=0.5-3$.  We show that the same population exists in
large numbers at low redshifts ($z\approx0.03$) and are actually the
edge-on manifestation of low surface brightness disk galaxies.

\end{abstract}

\subjectheadings{galaxies:evolution, galaxies:general, galaxies:irregular,
galaxies:structure}

\section{Introduction}

The launching of the Hubble Space Telescope has for the first time
allowed astronomers to observe the morphology of galaxies at moderate
redshifts.  Coupled with deep redshift surveys, {\sl HST} imaging
provides a powerful tool for studying the evolution of galaxies, in
particular, the ubiquitous but poorly understood population of
``excess'' faint blue galaxies.  A recent paper by Cowie et al.\
(1996; hereafter CHS) reports that many of the blue galaxies in deep
F814W {\sl HST} images of two fields from the Hawaii Redshift Survey
appear to be very narrow, linear structures.  The galaxies tend to be very
straight, and have sizes between 2-3\arcsec\ in the long dimension and
are marginally resolved after deconvolution at 0.05-0.1\arcsec\ in the
narrow dimension.  CHS argue that the extreme ellipticity of
these galaxies and the lack of continuity between the linear ``chain''
galaxies with the rest of the galaxy population is incompatible with
their being edge-on disk galaxies.  They therefore treat these
galaxies as a distinct morphological class of {\sl intrinsically}
linear galaxies which they refer to as ``chain'' galaxies.  Two of the
chain galaxies are emission line galaxies at redshifts of
$z\approx0.5$, another is argued to be at $z\approx1.4$ based upon one
strong emission line, and a fourth is suggested to be at $z=2.4$ based
on interpreting a break in the continuum at $4000$\AA\ as the onset of
intergalactic Ly-$\alpha$ forest absorption.  On the basis of the
emission lines and blue colors, coupled with the short lifetime
expected for intrinsically linear structures, CHS
argue that their new morphological class of chain galaxies are not
only star-forming, but are in the process of formation (see also
Ogorodnikov 1967).

The conclusions of CHS that chain galaxies represent a new
morphological class of linear galaxies in the process of forming at
$z=0.5-3$ is based on the following observations: 1) the extreme
ellipticities 2) the lack of continuity between the chain galaxies and
the rest of the galaxy population and 3) the lack of any low redshift
counterparts.  In this paper we will first show that the extreme
ellipticities are not incompatible with chain galaxies being
intrinsically disky systems viewed edge-on.  We will then use data from
an ongoing ground-based redshift survey of low surface brightness
galaxies to show that there are large numbers of galaxies with similar
morphology at very low redshifts ($z\approx0.03$) and that they join
seamlessly onto the population of edge-on normal galaxies.
Furthermore, we will show that the low redshift galaxies with the
``chain'' morphology are consistent with their being the edge-on
manifestation of face-on low surface brightness galaxies (LSBs) found
in the same survey.

\section{Extreme Ellipticities \& Morphology}	\label{ellipticities}

It is on the basis of their very large axial ratios that CHS argue
that the chain galaxies cannot be intrinsically disk systems.  The 21
chain galaxies observed in the CHS images have a mean axial ratio of
$4.7\pm2.3$ at the 20\% of peak isophotal level in the undeconvolved
surface brightness profile, with a maximum observed axial ratio of
9.5.  When the images are deconvolved with the point spread function
(FWHM$=0.19\arcsec-0.22\arcsec$), CHS find transverse widths of
$0.05-0.23\arcsec$, which are $0.25-1.15$ times the full width at half
maximum of the HST point spread function.  Assuming that the observed
axis length is the quadrature sum of the intrinsic axis length and the
width of the point spread function at half maximum, we can calculate
the distribution of intrinsic axis ratios from the undeconvolved axial
ratios published by CHS.  These are plotted in Figure
\ref{axialratiofig} as a function of their observed axial ratios, with
the symbols coding the deconvolved transverse widths of the galaxies.
While the bulk of the chain galaxies have deconvolved axial ratios of
between 2 and 14, there are some galaxies with deconvolved axial
ratios extending up to 33.  Note, however, that all of the very large
axial ratios are associated with extremely small deconvolved minor
axis lengths; all of the galaxies with $a/b>20$ have deconvolved minor
axis lengths which are significantly smaller than a single WFPC2
pixel.  Therefore, the largest axial ratios are by far the most
uncertain.

We wish to show that such extreme axial ratios are not necessarily
inconsistent with disk galaxies viewed edge-on.  Rather than invoking
theoretical models, we can resort to observations of extremely
thin edge-on galaxies at low redshift to demonstrate that 
galaxies with the ``chain'' morphology do not necessarily require
a new morphological class.

In the past twenty-five years there has been a sporadic commentary on
the existence of very elongated edge-on galaxies.
Vorontsov-Velyaminov (1967) and de Vaucouleurs (1974) both give
examples of galaxies with axial ratios which are larger than 10:1.
The published photographs of these flat galaxies all show the knotty
structure that CHS identify with the ``chain'' galaxies.  Most
illuminating are the additional examples from Vorontsov-Velyaminov
(1967) which have both a extremely flat, low surface brightness disk
($a/b>20$) {\sl and} a prominent bulge; this provides uncontrovertible
evidence that there are true disks at low redshift with the extreme
axial ratios observed by CHS.  Furthermore, a dynamical study of
extremely thin galaxies by Goad \& Roberts (1981) show that the
galaxies are rotating around their minor axis, as would be expected if
they were simply edge-on disks.

More recently, Karachentzev et al.\ (1993) have compiled a
catalog of edge-on galaxies (the Flat Galaxy Catalog) which have
$a>40\arcsec$ and $a/b \ge 7$ on the Palomar Observatory Sky Survey
and ESO/SERC survey plates.  An analysis of the catalog by Kudrya et
al.\ (1994) shows that there are extremely flat low-redshift galaxies
which have $a/b\approx22$, flatter than 80\% of the CHS chain
galaxies.  Furthermore, using the distribution of observed axial
ratios to determine the maximum disk flattening (i.e.\ to correct for
inclination effects), they find that the galaxies become progressively
thinner with later Hubble types, with the limiting axial ratio varying
from $a/b_{max}=14.1$ for Sb galaxies to $a/b_{max}=27.0$ for Sd's.
Thus, there {\sl are} edge-on disks which are as thin as the galaxies
in the CHS data, particularly among the later Hubble types.

\section{``Chain'' Galaxies at Low Redshift}		\label{lowz}

The steady increase in disk ellipticity with increasing Hubble type
observed in the Flat Galaxy Catalog is also accompanied both by a
decrease in the mean surface brightness and by an increase in the
fraction of galaxies with pronounced asymmetry (from 17\% for Sb
galaxies to 83\% for Sdm galaxies) (Karachentzev et al.\ 1993, Guthrie
1992).  These trends are also manifested in the high-redshift chain
galaxies, where the thinnest galaxies are also marked by lumpy
morphologies embedded in an extended, linear low surface brightness
component.

It is not unreasonable to assume that the increase in axial ratios,
decrease in surface brightness, and increase in asymmetry would
continue ever further along the Hubble sequence into the low surface
brightness Sm/Im classes.  There are many examples of Im and Sm
galaxies which, when viewed face on, appear to be disks with only one
or two HII regions superimposed.  If such galaxies were viewed edge
on, and particularly if they continue the trend for late-type galaxies
to have progressively thinner disks, then they would likely be
observed to have the chain morphology.  Most importantly, Marzke et
al.\ (1994) have shown that Sm/Im galaxies begin to dominate the
luminosity function fainter than $M^*-1$, and thus the high surface
density of galaxies with the chain morphology at intermediate
redshifts may not be surprising, if the galaxies are primarily more
than one magnitude fainter than $M^*$.  The three chain galaxies in CHS
with well determined redshifts do have absolute
$K$ magnitudes between 1 and 4 magnitudes fainter than the value of $M^*$
reported by Mobasher et al.\ (1993); however, more redshifts are
needed before any conclusions about the expected surface density can
be made.

Based on the above trends, we believe that edge-on Sm and Im galaxies
are the true counterparts to the chain galaxies observed at
intermediate redshift.  To establish this link, however, we must first
identify low redshift galaxies which also share the chain morphology,
and then we must show the continuity between this population and the
rest of the Sm/Im galaxy population.  The most notable feature of the
Sm and Im classes is their very low surface brightness (with the
exception of the fraction of actively starbursting galaxies such
as NGC4449) and as such, the bulgeless low surface brightness galaxies
(LSBs) being found in recent surveys(Impey, Bothun, \& Malin 1988,
Bothun, Impey, \& Malin 1991, Schombert et al. 1992, Schombert \&
Bothun 1988, Irwin, Davies, Disney, \& Phillipps 1990, Turner,
Phillipps, Davies, \& Disney 1993, de Jong 1995, Dalcanton 1995 and
references therein) are best viewed as extreme members of this Hubble
type (Dalcanton et al.\ 1996).  Morphologically, many of the LSB
images published by Schombert et al.\ (1992) are indistinguishable
from Im galaxies observed in Virgo (Sandage \& Binggeli 1984),
although the Schombert et al.\ (1992) galaxies tend to have larger physical
sizes.  Thus, samples of nearby LSB galaxies are likely places to find
low redshift galaxies with the chain morphology observed at
intermediate redshifts.

Further support for the LSB-chain galaxy connection comes from
a dynamical study of low redshift ``superthin'' galaxies
by Goad \& Roberts (1981).  After measuring rotation curves for four
galaxies with axial ratios up to 20:1, as well as for two comparison
galaxies, Goad \& Roberts found that the ``superthin'' galaxies
have very shallow velocity gradients in their centers.  The slow
rise towards the halo-dominated, flat portion of the rotation curve
suggests that the disk of the superthin galaxies contributes very
little mass to the inner parts of the galaxies.  As discussed in
Dalcanton et al.\ (1996), this is a general property expected for
the low surface density disks of LSBs.

\subsection{The Las Campanas Redshift Survey of LSBs}	\label{survey}

We have recently completed an extension to the Las Campanas Redshift
Survey (LCRS; Shectman et al.\ 1992) which is designed to measure the
contribution of galaxies with intrinsically low surface brightness to
the local galaxy luminosity function.  The new survey, the Low Surface
Brightness Extension (LSBX), used 14.5 square degrees of the original
$r$ band survey data from the LCRS to identify a sample of 672
galaxies with faint aperture magnitudes (up to 1.5 magnitudes fainter
than the aperture magnitudes in the LCRS) but large angular scale
lengths; this procedure aims to identify intrinsically low surface
brightness galaxies within a volume comparable to the volume probed by
the LCRS, in spite of the fainter magnitude limit.  The observed
galaxies were chosen randomly from among all galaxies satisfying the
selection criteria.  The multi-fiber spectrograph on the Dupont 2.5m
at Las Campanas (Shectman 1993) was used with a resolution of 5\AA\ to
measure redshifts to the target galaxies in three different fields,
with an average exposure of 8 hours per field.  After sky subtraction,
cross-correlation with A and F star templates, and identification of
emission lines, we successfully measured redshifts for 91\% of the
galaxies in the sample; the median redshift of the sample is $z=0.11$,
and 90\% of the galaxies lie at redshifts less than 0.25.  The details
of the data acquisition and reduction will be discussed in Dalcanton
\& Shectman (1996).

Because the resulting sample, when combined with the original LCRS,
spans a range of central surface brightness of over a factor of 40, it
provides a nearly ideal sample for 1) identifying galaxies with the
chain morphology at low redshifts and 2) demonstrating the continuity
of the chain morphology with normal galaxies.  For the purposes of
this paper, we will be working with a volume limited subsample,
consisting of the 75 galaxies in our sample with recessional
velocities between 10,500$\kmsec$ and 15,500$\kmsec$
($z_{median}=0.038$, and 90\% of the galaxies are within $1000\kmsec$
of the median velocity); the velocity range was chosen to fully
encompass a significant ``wall'' in the galaxy distribution.  The
subsample contains galaxies with isophotal $r$-band
magnitudes between 14.5 and 19.

\subsection{Continuity With Edge-On Normal Galaxies}	\label{continuity}

To show that there are low-redshift galaxies with the chain morphology
and that they form a continuum with high-surface brightness edge-on
normal galaxies, we selected galaxies from the volume limited
subsample which have ellipticities ($\epsilon\equiv1-b/a$) greater
than 0.6.  Figure \ref{edgeonfig} shows the eight galaxies, the five
faintest of which are from the low surface brightness extension,
arranged in order of decreasing mean surface brightness.  First, note
that within the limits of the ground based resolution, the low surface
brightness galaxies are close morphological analogs of the chain
galaxies observed at moderate redshift in HST imaging.  For example,
the five lowest surface brightness galaxies in Figure \ref{edgeonfig}
are remarkably similar to CHS's ``chain'' galaxies number 23, 14, 3,
9, and 4, including hints of the knots seen in the deeper,
high-resolution HST imaging.  Second, note that there is a natural
progression from edge-on normal galaxies, which have both a
recognizable bulge component and a high surface brightness disk,
through galaxies which have a high surface brightness disk but no
visible bulge and finally to highly elongated galaxies with very low
disk surface brightnesses.  This suggests that the extreme thin, low
surface brightness galaxies at the end of this sequence exist in
continuity with the population of normal disk galaxies viewed edge on.

To put this impression on a more quantitative footing, the axial
ratios were measured at the isophote corresponding to 20\% of the peak
surface brightness, following the method used by CHS, and are plotted
in Figure \ref{axialratioplot} as a function of their central surface
brightness\footnote{after including an optically-thin correction to
the observed peak surface brightness --
$\Delta\mu_0=2.5\lg{(\frac{a/b}{N})}$, where $N$ is the number of
exponential scale lengths corresponding to $a/2$.  $N=2.4$ for $a$
measured at the 20\% of peak isophotal surface brightness}.  All but
three of the twenty-one ``chain'' galaxies identified by CHS have
unconvolved axial ratios smaller than the largest axial ratio in our
small sample, and the axial ratios of the remaining three are only
20\% larger.  Thus, in the unconvolved data, there is very little
difference in the range of ellipticities of our sample and the CHS
sample.  Two of the eight galaxies in our sample are unresolved in the
minor axis, and thus their measured axial ratios are lower limits to
their true axial ratios.  Furthermore, the percentage of unresolved
galaxies (25\%) is the same as in the CHS data (Figure
\ref{axialratiofig}).  Therefore, the very thin structures being
uncovered in deep HST imaging have structural analogs at low redshift,
and are therefore not necessarily a manifestation of a unique state of
galaxy formation at moderate to high redshifts.

\subsection{Continuity With Face-On LSBs}		\label{faceon}

To show that the thin faint galaxies in Figure \ref{edgeonfig} are
actually disks, we must not only show their continuity with the disks
of normal galaxies; we must find their face-on low surface brightness
counterparts and show that the population as a whole is consistent
with thin disks seen from random viewing angles.  We show in Figure
\ref{faceonfig} all galaxies in the volume limited subsample with a
central $r$-band surface brightness fainter than 21.3 (roughly 0.5
magnitudes below the Freeman (1970) value, assuming $<B-r>\sim0.8$ (de
Blok et al.\ 1995)), and
in Figure \ref{inclinationplot}, we show their cumulative distribution
of ellipticities (again measured as in CHS).  The lightly
shaded region corresponds to the range of ellipticities which with a
galaxy having the median major axis length would be unresolved in the
minor axis, and the heavily shaded region marks the range of
ellipticities which with a galaxy whose major axis length was in the
top 95\% would be unresolved in the minor axis.  The measured
ellipticities of galaxies falling within this region are therefore
likely to underestimate the true ellipticities, and thus very little
interpretation can come from the detailed distribution of
ellipticities in this regime, beyond noting that the distribution is
continuous.  This again suggests that galaxies at low redshift with
the ``chain'' morphology are {\sl not} a disjoint population; they
form a continuous distribution with face-on LSB galaxies, in addition
to being morpholgically continuous with edge-on normal galaxies.  In
fact, the distribution of ellipticities outside of the heavily shaded
region is indistinguishable from what would be expected for a single
population of perfectly thin, circular disk galaxies viewed at random
inclinations (the straight diagonal line in Figure
\ref{inclinationplot}, which has only a 50\% chance of not being
representative of the distribution of ellipticities, by the
Kolmogorov-Smirnov test).  Clearly, there is nothing to suggest that
the very thin galaxies in Figures \ref{edgeonfig} \& \ref{faceonfig},
as well as in deep HST images, are anything more than low surface
brightness {\sl disk} galaxies viewed edge on.

Although the edge-on circular disk model provides a statistically
adequate fit to the distribution of ellipticities, there are
indications of a possible paucity of face-on galaxies.  This may
reflect an intrinsic non-cicularity of LSB disks.  We note in passing
that a better fit to the ellipticity distribution results from adding
the assumption that the disks are not perfectly circular, and have an
intrinsic ellipticity of $\epsilon_0=0.27$ (the curved line in Figure
\ref{inclinationplot}, calculated from Monte Carlo simulations of the
observed ellipticity of intrinsically noncircular disks; see Rix \&
Zaritsky 1995). Note that few, if any, of the galaxies in Figure
\ref{faceonfig} appear to be circularly symmetric galaxies viewed from
an angle; this is not surprising, given that vast majority of chain
galaxies in CHS are likewise asymmetric.

\section{Colors, Magnitudes, and Surface Brightnesses}		\label{colors}

Low redshift LSBs are among the bluest galaxies known (McGaugh \&
Bothun 1994, de Blok et al.\ 1995, Knezek 1993, de Jong 1995).
Similar claims are made for the chain galaxies published by CHS.
Although the almost total lack of redshifts for the chain galaxies
makes a direct comparison between chain and LSB colors somewhat
difficult, with some reasonable assumptions we can estimate
appropriate $k$-corrections for the chain galaxies.  
We first assume that the blue colors suggest that the CHS galaxies
have spectral energy distributions characteristic of irregular
galaxies.  Secondly we assume that the bulk of the chain galaxies
lie between $z=0.5$ and $z=1.5$, which encompasses the range spanned
by the three securely measured redshifts in the chain galaxy sample.
Adopting the $k$-corrections for Im galaxies published in Cowie et
al.\ (1994), the $B-K$ and $B-I$ $k$-corrections are therefore in the
ranges $1.2-1.4$ and $-0.1-0.4$, respectively.  The median $B-K$ color
for the CHS chain galaxies is 4.1, and their median $B-I$ color is
1.5; these correct to $\langle B-K\rangle\approx2.8$ and $1.1\lta
\langle B-I\rangle \lta 1.6$ at zero redshift.

We can compare the $k$-corrected chain galaxy colors to the published
$B-I$ colors of LSBs studied by de Blok et al.\ (1995) and by McGaugh
\& Bothun (1994), and to the $B-K$ colors of the face-on disk galaxies
studied by de Jong (1996).  The two former were drawn from the
Schombert et al.\ (1992) catalog and the Uppsala General Catalog (UGC;
Nilson 1973), and the latter were drawn entirely from the UGC.  The
LSBs in the de Blok and the McGaugh \& Bothun catalogs catalog have
$1<B-I<1.3$ and $1.1<B-I<2.3$ respectively, and the late type ($T>6$)
galaxies from de Jong (1996) have $B-K\approx 2.8$.  Therefore, the
range of colors spanned by the published LSB samples is completely
consistent with the blue colors of the CHS chain galaxy samples

We wish to make the additional point that the properties of the CHS chain
galaxies are included in the range of
surface brightness and absolute magnitude spanned by the low redshift
galaxies in Figure \ref{edgeonfig}.  The three CHS chain galaxies with secure redshifts have
$k$-corrected absolute $B$ magnitudes between 2.1 magnitudes below
$M_*$ and 1 magnitude above $M_*$ (using $M_*$ as calculated by
Efstathiou et al.\ 1988), while the low redshift galaxies in Figure
\ref{edgeonfig} have absolute $r$ magnitudes between 3.1 and 0.3
magnitudes below $M_*$ (using $M_*$ from Lin et al.\ 1996).  The
linear sizes of the galaxies in both samples suggest exponential disk
scale lengths of $1-4\,h_{75}^{-1}\kpc$ (given that the major axis
lengths were measured at 20\% of the peak surface brightness of an
edge-on disk).  The combination of similar absolute magnitudes
and similar scale lengths immediately suggest that the LSBs in
Figure \ref{edgeonfig} and the CHS chain galaxies have comparable
surface brightnesses.  This conclusion also holds if one
compares the mean surface brightnesses for each individual galaxy
in the two samples, assuming $B-r\approx0.8$ (de Blok et al.\ 1995).

\section{Conclusions}					\label{conclusions}

We have shown that the apparently unfamiliar population of very thin
galaxies revealed in deep HST imaging do in fact have low redshift
counterparts among the population of low surface brightness disk
galaxies, and thus, that there is no immediate need to invoke any
evolutionary scenario to explain their appearance at moderate
redshifts.  We have shown examples of such very thin galaxies in
existing catalogs of nearby galaxies, and shown the statistics of the
ellipticity distributions of low-redshift LSB galaxies are consistent
with the low redshift ``chain'' galaxies being edge-on manifestations
of disks.  This suggests that, at low redshift, such thin systems are
{\sl not} necessarily intrinsically linear (as opposed to disky),
bringing into question the assumption that equally thin systems at
high redshift {\sl cannot} be disks (CHS).  Our conclusion that the
thin ``chain'' galaxies observed at moderate to high redshifts are
analogs of low surface brightness disks is supported by both their
morphologies and by their colors as well.  LSBs at low redshift are
intrinsically asymmetric (Figure \ref{inclinationplot}), often with
one or two bright knots superimposed upon a diffuse low surface
brightness disk (Figure \ref{faceonfig}); such systems, when viewed
edge-on, would be consistent with the lumpy structure of some of the
chain galaxies.  The moderate redshift chain galaxy / low
redshift edge-on LSB class may be
viewed simply as a subset of the link between the faint blue galaxies
and nearby LSBs posited by Ferguson \& McGaugh (1995).

\acknowledgements

It is a pleasure to thank the members of the LCRS team for allowing
use of their imaging data in Figure \ref{edgeonfig} and in creating
the sample for the LSBX.  The referee, Greg Bothun, is thanked for
extremely helpful suggestions, as are Gus Oemler, Mauro Giavalisco,
and John Mulchaey for reading early drafts of the manuscript.

\vfill
\clearpage


\vfill
\clearpage

\figcaption[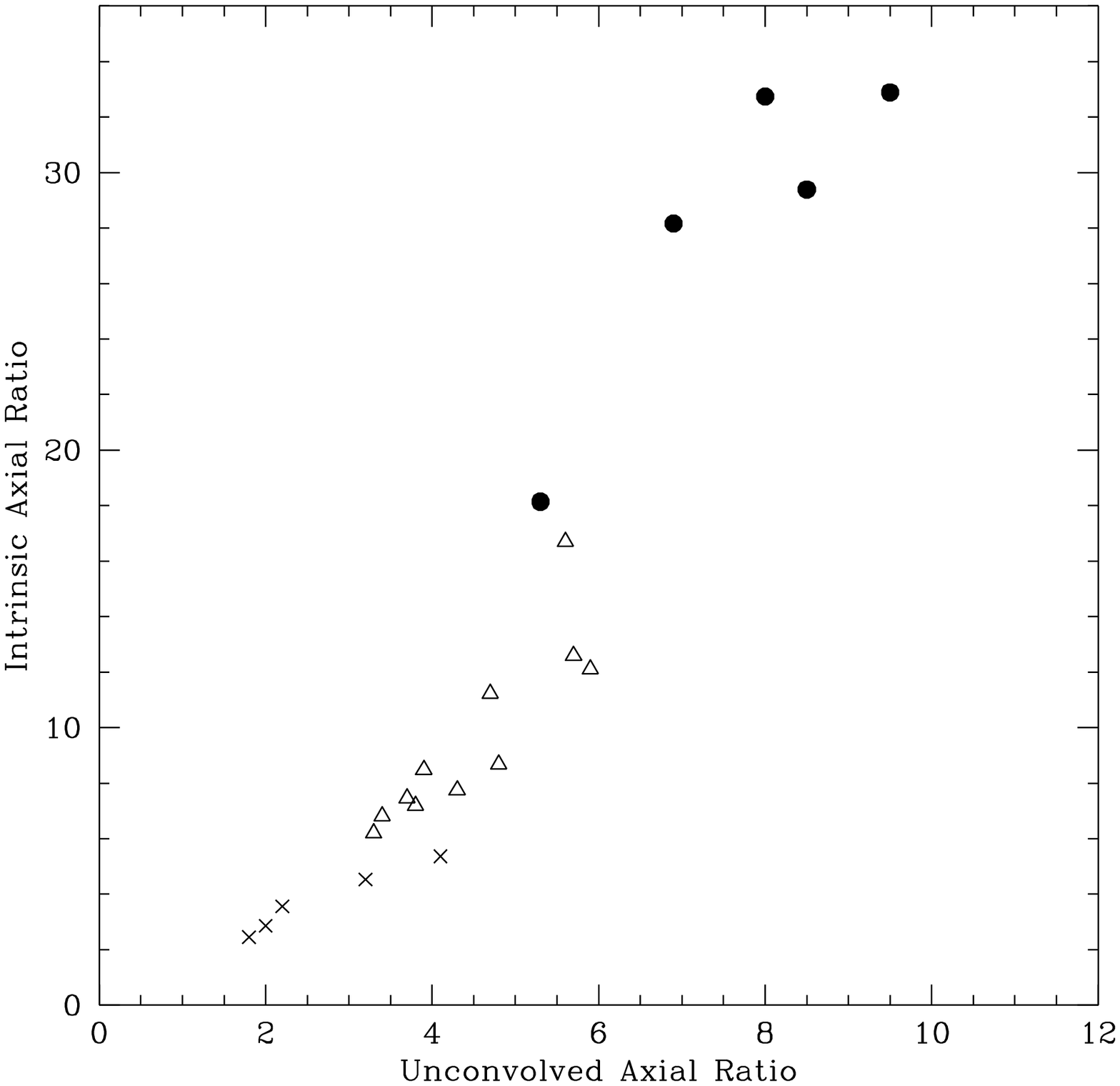]{Deconvolved axial ratios as a function of
measured axial ratio for the 21 Cowie et al.\ (1996) chain galaxies.
The deconvolved axis ratios are estimated using
$b_{observed}^2=b_{intrinsic}^2+\sigma^2$, where $\sigma$ is the width
of the point spread function at half maximum and $b$ is the minor axis
length, and assuming that $a_{observed}\approx a_{intrinsic}$.
Galaxies whose deconvovled minor axis lengths are less than $\sigma/3$
are marked with solid circles, between $\sigma/3$ and $2\sigma/3$ are
marked with open triangles, and all others are marked with crosses.
\label{axialratiofig}}

\figcaption[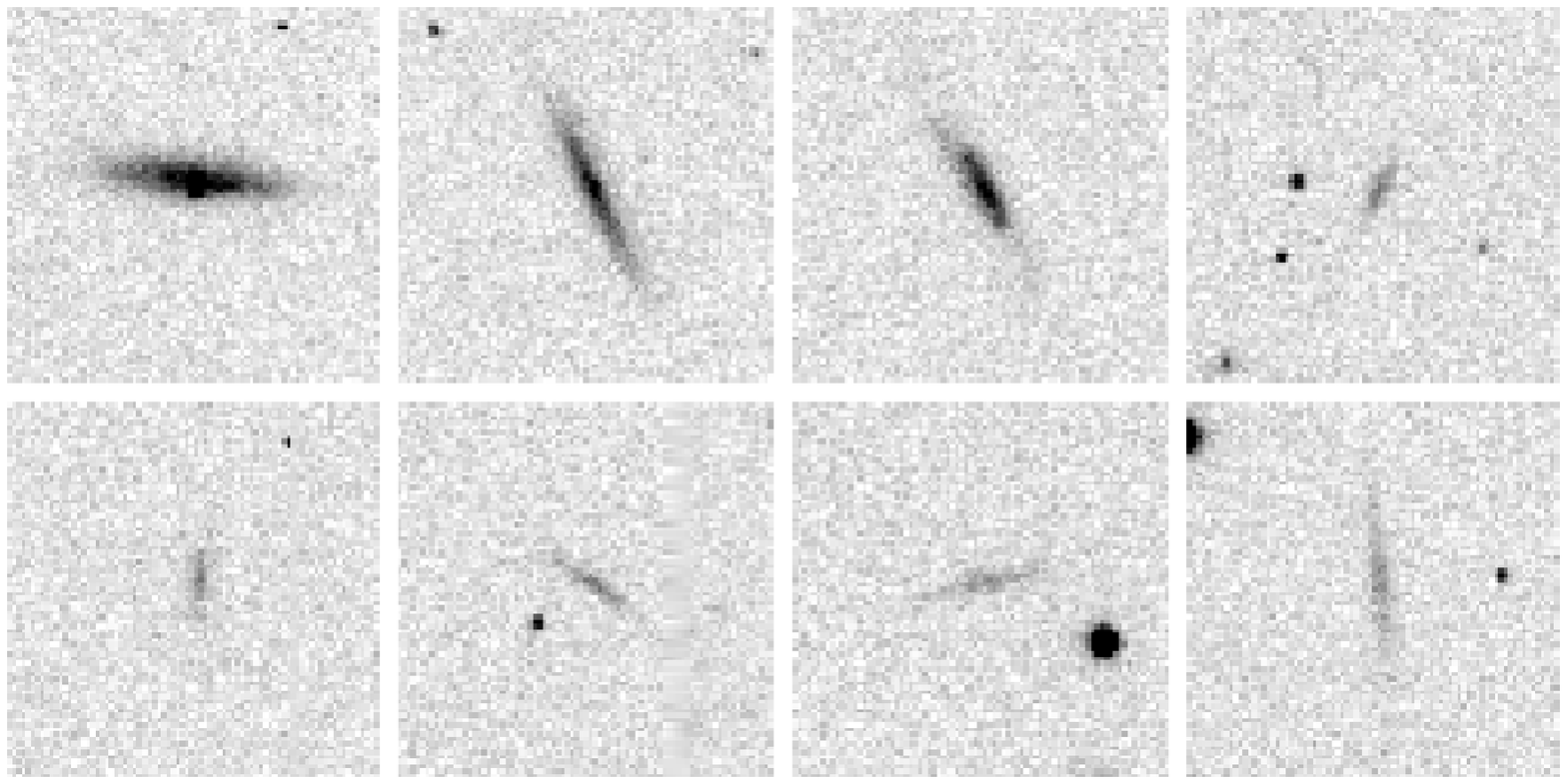]{ Edge-on galaxies ($\epsilon>0.6$) from the
volume limited $z\approx0.04$ LCRS+LSBX subsample, arranged in order
of decreasing central surface brightness.  Their distribution of axial
ratios as a function of central surface brightness is plotted in
Figure \ref{axialratioplot}.
\label{edgeonfig}}

\figcaption[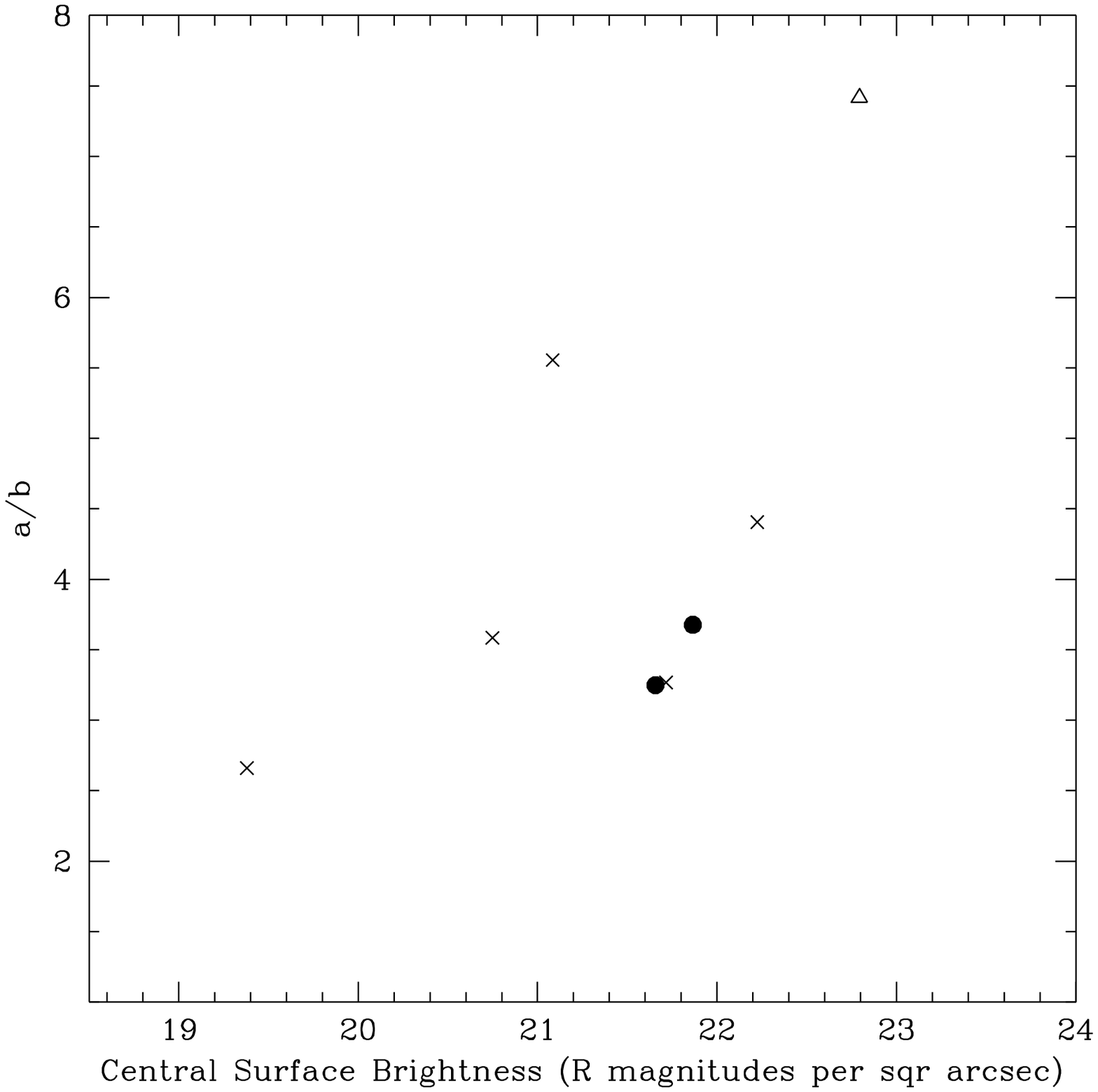]{ Axial ratio vs central surface brightness for
flattened galaxies in the volume limited $z\approx0.04$ LCRS+LCRX
subsample (Figure \ref{edgeonfig}).  The ellipticity is measured at
20\% of the peak surface brightness, analagous to CHS.  Galaxies whose
minor axis length are less than 33\% larger than the seeing are marked
with solid circles, between 33\% and 66\% are marked with triangles,
and all others are marked with crosses.  Note that the range of
observed axial ratios is almost identical to those observed in the
deep HST imaging (Figure \ref{axialratiofig}), although the galaxies
are at much lower redshift.
\label{axialratioplot}}

\figcaption[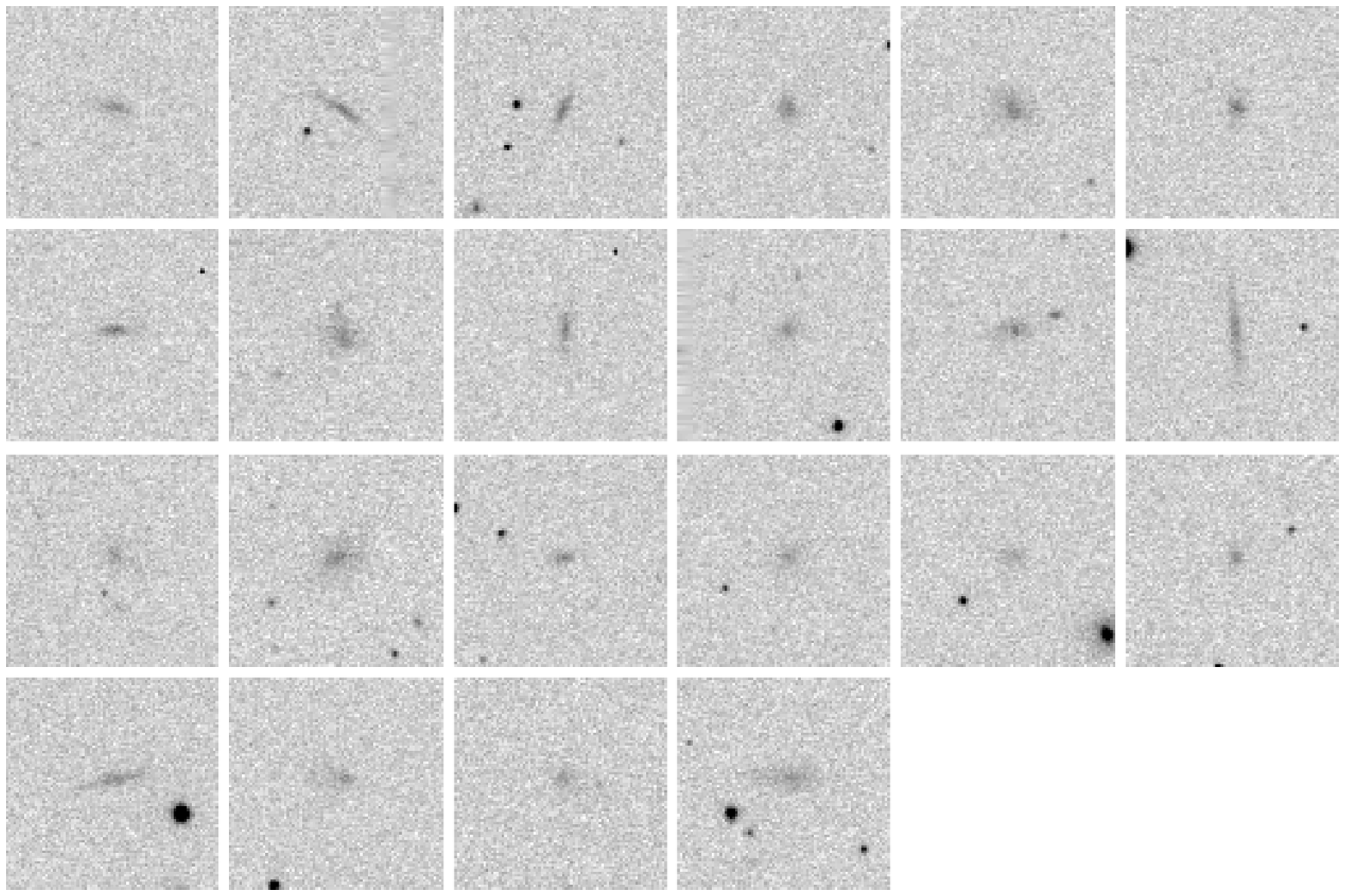]{ A volume limited sample of low surface
brightness galaxies ($\mu_0>21.3\,r\surfb$) from the volume limited
$z\approx0.04$ LCRS+LSBX subsample.  Note that some of the galaxies
have one or two assymetrically placed knots embedded within a lower
surface brightness halo; presumably, these systems would look very
similar to the knottier of the chain galaxies, if they were to be
viewed edge on at high resolution.  Their distribution of
ellipticities is plotted in Figure \ref{inclinationplot}.
\label{faceonfig}}

\figcaption[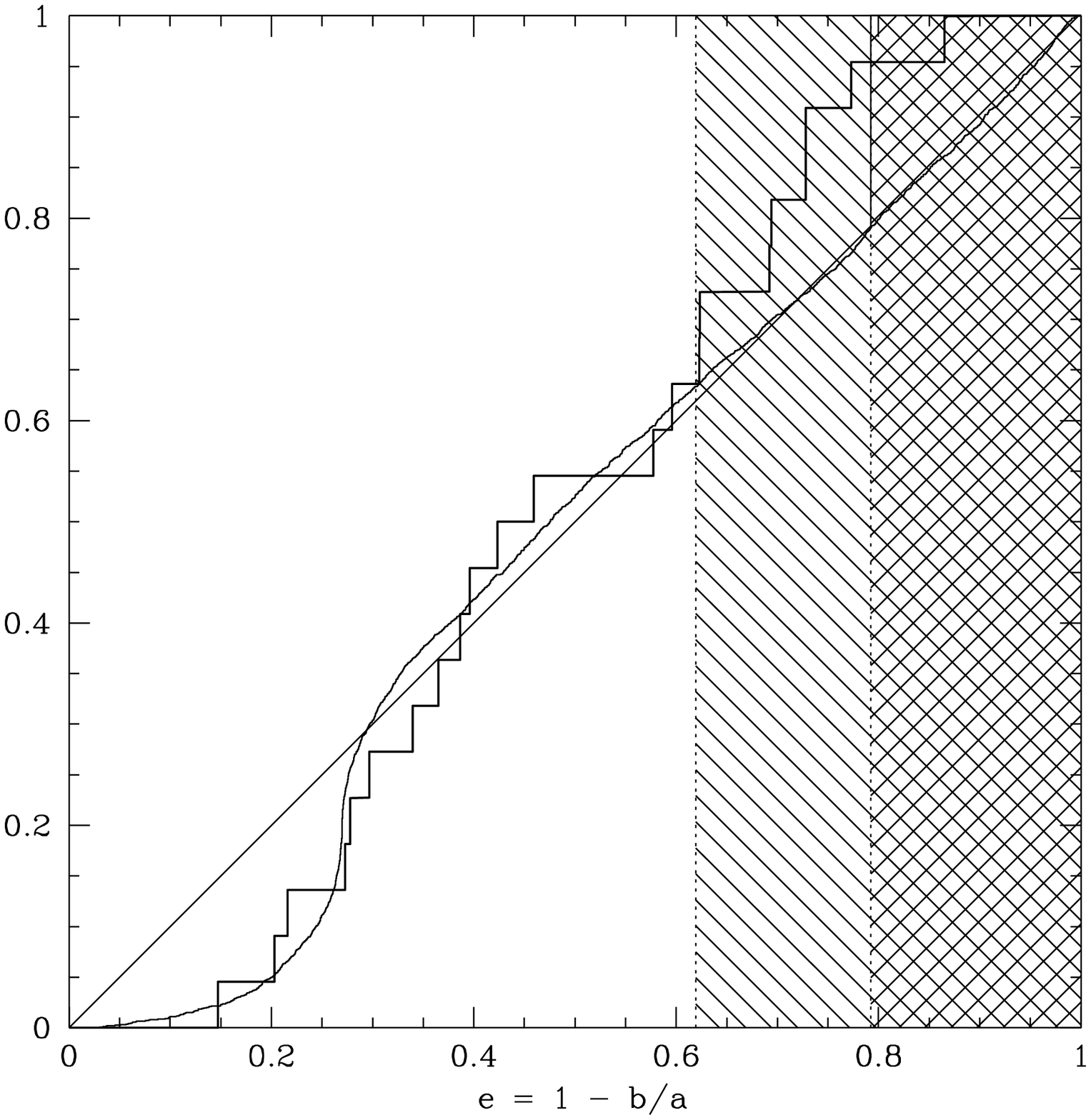]{ Cumulative distribution of ellipticities of
the LSB subsample from Figure \ref{faceonfig} (heavy stepped line),
compared to the predicted cumulative distribution of ellipticities for
perfectly flat circular disks (diagonal light line) and perfectly flat
elliptical disks with intrinsic ellipticity $\epsilon_0=0.27$ (curved
light line).  The median uncertainty in measured ellipticities is
$\pm0.1$.  To show how seeing limits the maximum observed ellipticity,
the light hashed region is where a galaxy with the median major axis
length would have a minor axis length that was less than 25\% larger
than the seeing width, and the dark shaded region is where a galaxy
whose major axis was larger than 95\% of the galaxies in the subsample
would have a minor axis length that was less than 25\% larger than the
seeing width; in these regions, the distribution of ellipticities is
expected to deviate from random.
\label{inclinationplot}}

\setlength{\topmargin}{-0.5in}

\thispagestyle{empty}
\begin{figure}[t]
\centerline{ \psfig{file=figure1.ps} }
\begin{flushright}{\bigskip\cap Figure 1}\end{flushright}
\end{figure}
\vfill
\clearpage

\thispagestyle{empty}
\begin{figure}[t]
\centerline{ \psfig{file=figure2.ps,height=2in} }
\begin{flushright}{\bigskip\cap Figure 2}\end{flushright}
\end{figure}

\thispagestyle{empty}
\begin{figure}[b]
\centerline{ \psfig{file=figure3.ps,height=4in} }
\begin{flushright}{\bigskip\cap Figure 3}\end{flushright}
\end{figure}
\vfill
\clearpage

\thispagestyle{empty}
\begin{figure}[t]
\centerline{ \psfig{file=figure4.ps,height=3in} }
\begin{flushright}{\bigskip\cap Figure 4}\end{flushright}
\end{figure}

\thispagestyle{empty}
\begin{figure}[b]
\centerline{ \psfig{file=figure5.ps,height=3.5in} }
\begin{flushright}{\bigskip\cap Figure 5}\end{flushright}
\end{figure}
\vfill
\clearpage
\end{document}